\documentclass[a4paper,twocolumn,letter]{aa}

\usepackage{graphicx}
\usepackage{txfonts}
\usepackage{lipsum}
\usepackage{lscape}             
\usepackage{placeins}           
\usepackage{hyperref}
\usepackage{wrapfig}

\usepackage{xcolor}

\usepackage{booktabs}
\usepackage{array}

\newcommand{\gaia}{\textit{Gaia}} %

\newcolumntype{L}[1]{>{\raggedright\arraybackslash}p{#1}}
\newcolumntype{C}[1]{>{\centering\arraybackslash}p{#1}}
\newcolumntype{R}[1]{>{\raggedleft\arraybackslash}p{#1}}

\title{A bird's eye view of stellar evolution through populations of variable stars in Galactic open clusters\thanks{Tables \ref{tab:catalog} and \ref{tab:catalog-photometry} are available at the CDS via \url{https://cdsweb.u-strasbg.fr/cgi-bin/qcat?J/A+A/700/L13}}}

\titlerunning{Variable stars in Galactic open clusters}
\author{
Richard~I. Anderson\inst{1}\thanks{\email{richard.anderson@epfl.ch}}
\and
Emily L. Hunt\inst{2,3}\thanks{Both authors contributed equally. \email{emhunt@mpia.de}}
}

 \institute{
 Institute of Physics, \'Ecole Polytechnique F\'ed\'erale de Lausanne (EPFL), Observatoire de Sauverny, 1290 Versoix, Switzerland \and 
 Max-Planck-Institut f\"ur Astronomie, K\"onigstuhl 17, 69117 Heidelberg, Germany
 \and
 Landessternwarte, Zentrum f\"ur Astronomie der Universit\"at Heidelberg, K\"onigstuhl 12, 69117 Heidelberg, Germany
  }

\date{Received 11 April 2025; accepted 10 July 2025} 

\abstract{Both star clusters and variable stars are sensitive laboratories of stellar astrophysics and evolution: cluster member stars provide context for interpreting cluster populations, whereas variability reveals the nature of individual stellar systems. The European Space Agency's \gaia\ mission has revolutionized the census of star clusters in the Milky Way, while simultaneously providing an unprecedented homogeneous all-sky catalog of variable stars. Here, we leverage the third \gaia\ data release to obtain an empirical bird's eye view of stellar evolution based on 34\,760 variable stars residing in 1\,192 Galactic open clusters (OCs) containing $173\,294$ members (variable member fraction $20.0\%$). Using precise OC distances, dereddened magnitudes, and consistently determined ages, we a) pinpointed regions of pulsational instability across the color-absolute magnitude diagram (CaMD); b) traced the occurrence rate of variables as a function of age, and c) considered the evolution of rotation periods and photometric activity (gyrochronology). The occurrence of pulsating stars can serve as a model- and reddening-independent age estimator. 
Our results underline that jointly considering stellar variability and OC membership enables a plethora of further applications, such as age dating or dereddening OCs based on expected CaMD locations of variable stars. Upcoming \gaia\ data releases and the Vera C. Rubin Observatory will vastly increase the extent to which the details of variable stars in OCs can empirically unravel the astrophysics and evolution of stellar populations.}


\begin{document}

\maketitle\

\section{Introduction}

Open clusters (OCs) are fundamental laboratories of stellar evolution. They are typically considered as simple stellar populations that trace stellar evolution as a function of age and metallicity,  and their color-magnitude diagrams (CMDs) provide context for interpreting the nature of their individual member stars. Although the majority of OCs that form initially are dispersed on timescales of $\sim${}10\,Myr \citep{Lada2003}, the ages of observed OCs span several orders of magnitude, up to a few Gyr \citep[e.g.,][]{CantatGaudinReview}. 
The \gaia\ mission \citep{GaiaCollaborationPrusti_2016} has revolutionized the Galactic OC census \citep{GDR1_clusters} with ever-improving completeness and purity \citep[e.g.,][]{Cantat-GaudinAnders_2020,HuntReffert_2023}.

Variability constitutes a powerful laboratory for stellar astrophysics, since it unravels the structure and nature of individual objects 
\citep[e.g.,][]{AertsBook} 
as well as stellar populations.
Such information can complement constraints provided by CMDs. Large time-resolved surveys, such as the Optical Gravitational Lensing Experiment \citep{OGLE_UdalskiSzymanski_2008} and VISTA Variables in the V\'ia L\'actea \citep{VVVdescription}, among others,
have revolutionized research on variable star populations. In particular, \gaia\ data release 3 (GDR3) has provided a homogeneous all-sky census of $9.5$\,million variable stars \citep{GaiaDR3_summary,GaiaDR3_variability_eyer}.

Variable stars residing in OCs combine the two complementary types of constraint and have a long history of study, with the first variable star members of OCs (classical Cepheids) already identified  70 years ago \citep[cf. also \citealt{SawyerHogg1959review}]{Irwin1955}. The WEBDA database\footnote{\url{https://webda.physics.muni.cz/}} collected detailed information on OCs and variable stars residing therein \citep[last update: 2013]{Zejda2012}. However, \gaia's capacity to identify large numbers of bona fide variable stars residing in Galactic OCs \citep{Anderson2013} has yet to be exploited more generally.
This Letter showcases the power of jointly considering these two concepts in the form of populations of variable stars in OCs. The high-quality and homogeneous GDR3 data thus create a bird's eye view of stellar evolution and illustrate how variable star populations evolve over time.

\section{Data and methods}\label{sec:data_methods}
We created a high-quality catalog of bona fide variable star members in high-quality OCs from \citet[hereafter: HR24]{HuntReffert_2024}. 
We adopted HR24's OC ages, $\log{t}$, and distances, $d$. We limited  the sample to $d\le 2$\,kpc, astrometric $S/N > 5$, median CMD quality $> 0.5$ \citep[following][]{HuntReffert_2023}, and total extinction $A_V < 1.5$. Bona fide OC member stars identified in HR24 with $G \leq 18$\,mag were considered to limit any issues related to larger photometric uncertainties.  We estimated the effective temperatures of members stars using PARSEC v1.2s isochrones \citep{BressanMarigo_2012} and dereddened all member stars using extinction relations for GDR3 photometric passbands \citep{AndersKhalatyan_2022} 
assuming a single $A_V$ value per OC and the \citet{SchlaflyMeisner_2016} extinction law. A systematic uncertainty of $\sigma_{A_V}\approx 0.1$\,mag (HR24) applies.

Variable OC member stars were identified in the GDR3 specific object studies (SOS) catalogs of long-period variables \citep[LPVs]{GaiaDR3_SOS_LPVs}, RR Lyrae stars \citep{GaiaDR3_SOS_RRLyrae}, Cepheids \citep{GaiaDR3_SOS_Cepheids}, eclipsing binaries \citep{GaiaDR3_SOS_ECL},  supervised classification \citep{GaiaDR3_variability_classifications}, and solar-like stars exhibiting rotational modulation \citep{GaiaDR3_SOS_Rotation}. We manually added the well-known OC Cepheid U~Sgr from the supervised classification. Pulsation periods were supplemented where available from SOS main sequence oscillators \citep{GaiaDR3_ms_oscillators}.
We inspected the color-absolute magnitude diagram (CaMD) positions of all variable stars in  Sect.\,\ref{sec:results} and found only five  misclassifications (see Appendix~\ref{app:misclassified_variables}).

\section{Results}\label{sec:results}

Our high-confidence catalog contains $34\,760$ variable stars across $1192$ OCs containing $173\,294$ members. The catalog is made available via the Centre de Donn\'ees de Strasbourg (CDS), including the dereddened photometry of OC members, cf. Tables~\ref{tab:catalog} and \ref{tab:catalog-photometry}. Table~\ref{tab:variables} lists its contents per variability class. Appendix\,\ref{app:example_clusters} shows examples of five particularly well-populated OCs. Generally speaking, variable star detectability depends on photometric amplitudes, $\mathcal{A}$, and precision, $\sigma_m$, in addition to the sampling, baseline, and more. Detectability thresholds of variables in OCs differ substantially because of large ranges of extinction-corrected absolute $G-$band magnitudes, $M_{G,0}$, and $d$ considered as well as the large diversity of variability types and $\mathcal{A}$, whose definition depends on the type of variations under study. We considered restricted apparent $G-$band magnitude, $m_G$, ranges in the figure comparisons to avoid saturation and/or high $\sigma_m$ as indicated in Appendix\,\ref{app:detectability}. Detailed investigations of variable star detectability across our OC sample will become possible with the fourth \gaia{} data release (GDR4).

\begin{figure}
\centering
\includegraphics[]{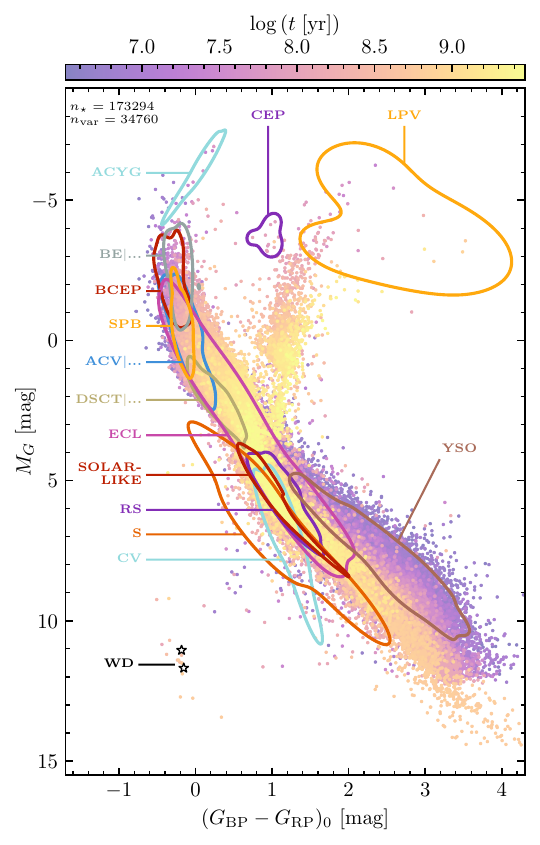}
\caption{ \gaia\ CaMD of variable OC member stars color-coded by age. Contours show regions occupied by specific variable types (at least three specimens), cf. Sect.\,\ref{sec:results:camd} and Table~\ref{tab:variables}. Two ZZ~Ceti pulsating white dwarfs in the Hyades are shown as black open asterisks.\label{fig:variable_hr}}
\end{figure}

\subsection{Variables across the CaMD}\label{sec:results:camd}

Figure~\ref{fig:variable_hr} shows the dereddened CaMD of our catalog with \gaia\ variability classifications \citep[cf.][]{GaiaDR3_variability_eyer}.
Locations occupied by specific classes are highlighted by contours defined by a kernel density estimation (KDE) algorithm and a 20\% density threshold. Additional illustrations are available in Appendix~\ref{app:camd_individual_stars}, including the location of individual variable stars and variable star CaMD tracks in Fig.~\ref{fig:variable_hr_with_points}.
Figure~\ref{fig:variable_hr} illustrates the rich diversity of variability across many phases of stellar evolution, complementing related CaMDs in \citet[their Figs.\,$3-7,$]{GaiaCollaborationEyer_DR2CaMD_2019} and in Appendix\,E of \citet{GaiaDR3_variability_classifications}.
Several pulsating star classes populate the upper main sequence, generally following well-known trends with luminosity \citep[e.g.,][]{Kurtz2022}. Beginning at the high-luminosity end, five $\alpha$ Cygni variables occupy a narrow region that seems to have evolved off of the main sequence. Below them, several variability types including $\beta$ Cephei, slowly pulsating B stars (SPBs), and different Be-type variables cover the entire upper main sequence as multiple overlapping variability regions. This overlap may be partially due to ambiguity in assigning variability types based on light curves alone. Further inspection of these overlapping CaMD regions may serve to understand and mitigate confounding factors of classification. Lower on the main sequence are RS Canum Venaticorum (RS CVn) stars and solar-like variables. Young stellar objects (YSOs) appear noticeably above the main sequence. 
Fourteen short-timescale variables (timescales tens of minutes to 1 d) are found in 13 OCs, slightly blueward of the main sequence between $2.7 \le M_{G,0} \le 11.1$\,mag. These stars are redder than most short-timescale variables \citep[cf. Fig.\,17 in][]{Roelens2018}.
LPVs occupy a large color range among the highest-luminosity giants, and classical Cepheid variables occupy the classical instability strip.

\subsection{Local variability fractions}\label{sec:results:localfrac}
Instability strips define specific regions on the Hertzsprung-Russell diagram where stars are unstable to pulsation due to their structure. However, recent studies of $\delta$~Scuti stars, classical Cepheids, and RR Lyrae stars have shown differing levels of impurity across the classical instability strip \citep{Murphy2015,Narloch2019,CruzReyes2024rrl}. This feature cannot currently be explained by stellar theory. In principle, populations of variable stars in OCs are well suited to extend such analyses to other variability types. 
However, \gaia's per-epoch photometric errors \citep[$2-5$\,mmag in the optimal $m_G$ range, cf.][]{GDR3_GAPS,MaizApellaniz2023} restrict its ability to detect small-amplitude variations. For example, $\delta$~Scuti star amplitudes detected by {\it Kepler} are $\sim 2$\,mmag and $\mathcal{A}<10$\,mmag for the vast majority \citep{Bowman2018}, whereas the lowest $\mathcal{A} = 9$\,mmag in our sample, cf. Table~\ref{tab:variables}. Similar limitations apply to, e.g., SX~Phoenicis, $\gamma$~Doradus, and rapidly oscillating chemically peculiar A (roAp) stars \citep[cf.][]{Catelan2015}. \gaia's temporal sampling further impacts the detectability.
Table~\ref{tab:variables} reports $\mathcal{F}_\mathrm{var}$, the fractions of stars detected as variables within each CaMD contour in Fig.\,\ref{fig:variable_hr}, which are not measures of purity as per the aforementioned limitations.
Among the extrinsic variables, the highest values of $\mathcal{F}_{\mathrm{var}}$ are found in the regions occupied by YSOs ($32\%$), solar-like ($22\%$), and RS~CVn variables ($10\%$). We find an incidence of $0.30 \pm 0.01\%$ eclipsing binaries,  which is much lower than reported using the continuously sampled and highly precise {\it Kepler} data \citep[$\sim 1.3\%$]{Kirk2016}, albeit over a larger mass range.

\begin{figure}[t]
\centering
\includegraphics[width=\columnwidth]{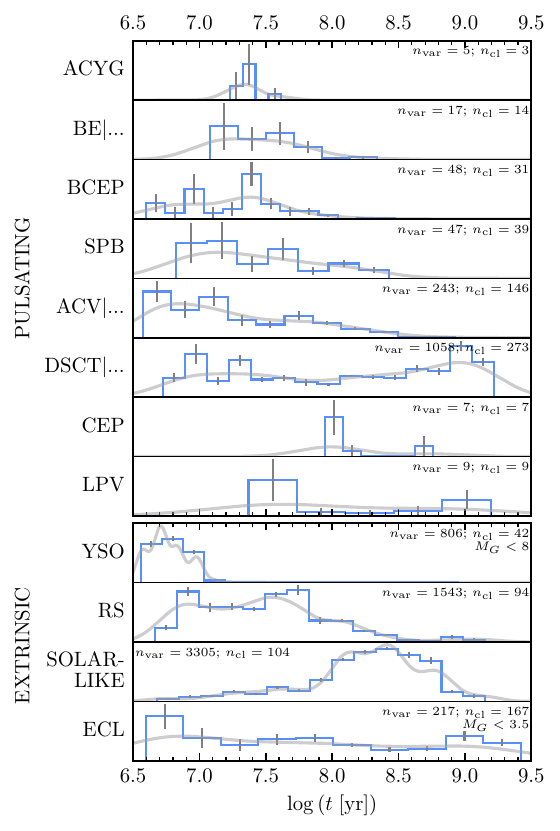}
\caption{Occurrence of variable stars in classes with at least five members against OC age. Sample restrictions have been applied to ensure consistent detectability limits for each class, cf. Appendix\,\ref{app:detectability}.  For YSOs and ECLs, an additional limit on $M_G$ has been applied as labeled. Labels on the left abbreviate  GDR3 variability classes, cf. Table~\ref{tab:variables}. The number of variables is normalized by the number of OC members with $4 \le M_G \le 5\,$[mag]. Dark gray error bars show Poisson uncertainties, while light gray lines show the KDE-smoothed distribution.
\label{fig:joy_division}}
\end{figure}

\subsection{Variability as a function of age}\label{sec:results:age}

Figure~\ref{fig:joy_division} illustrates the occurrence of variable star types against OC age, normalized by the number of member stars with $4 \le M_{G,0}\le 5\,$mag. Sample restrictions according to Appendix\,\ref{app:detectability} are applied to limit the effect of detectability differences.
The global variable star fraction as a function of age is shown in Appendix\,\ref{app:varfrac_age}. In general, younger OCs exhibit the most variable stars (cf. Fig.~\ref{fig:variable_member_fraction}) and OCs younger than $\sim${}100\,Myr exhibit the greatest diversity of variables. Examples of five OCs spanning ages from $5-500$\,Myr are shown in Fig.\,\ref{fig:example_clusters}.

Figure\,\ref{fig:joy_division} shows pulsating star classes occupying the upper main sequence increasingly disappear as the main sequence turn-off moves towards lower luminosity (and mass). The cleanest effect is seen for the chemically peculiar and rapidly oscillating A-stars (ACV|...). The DSCT|... group spans the OC age range rather uniformly. Poor statistics limit the ability to trace age dependencies for $\alpha$ Cygni stars, 
as well as classical Cepheids \citep[for a larger sample cf.][]{CruzReyes2023} and LPVs, both of which obey period-age relations \citep{Anderson2016rot,Trabucchi2022}.

Ninety percent of YSOs in our sample reside in OCs younger than 20\,Myr.
RS CVn stars are found predominantly in young OCs, although they persist to beyond $300$\,Myr. Above $100$\,Myr, a sudden decrease among RS~CVn is matched by a sudden increase in SOLAR\_LIKE types.
This echoes the observation in Sect.\,\ref{sec:results:gyrochronology} that activity signatures measured by excess photometric scatter dominate at younger ages and is consistent with both variability types being related to solar-like stars and magnetic activity  \citep{berdyugina_starspots_2005,GaiaDR3_variability_classifications}.

Eclipsing binaries do not exhibit a significant trend with age, although the KDE-smoothed distribution suggests a slight decrease with $\log{t}$, which would match the expectation that multiplicity fractions depend on mass \citep{OffnerMoe_2023}.

\subsection{Gyrochonology}\label{sec:results:gyrochronology}

\begin{figure}
\centering
\includegraphics[width=\columnwidth]{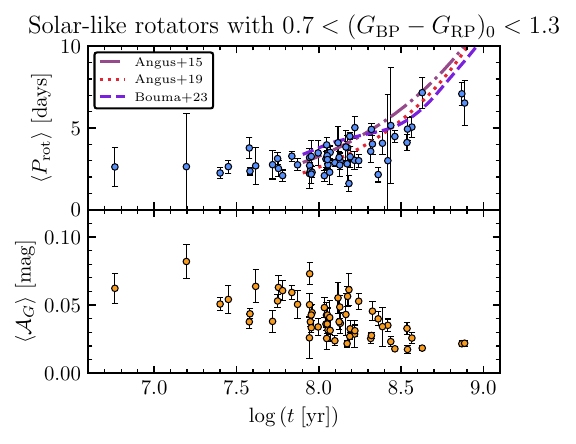}
\caption{\emph{Top:} 
Median rotation period, $\langle P_{\mathrm{rot}} \rangle$, of solar-like rotators in GDR3 against age for 58 OCs, cf. Sect.\,\ref{sec:results:gyrochronology}. 
Error bars show the standard error on the median per OC. Literature relations from \cite{AngusAigrain_2015}, \cite{AngusMorton_2019}, and \cite{BoumaPalumbo_2023} are shown as labeled. \emph{Bottom:} Same as the top, but for the per-OC median G-band activity index, $\langle \mathcal{A}_G \rangle$. \label{fig:gyrochronoloy}}
\end{figure}

Gyrochronology determines stellar ages based on activity indicators related to stellar rotation and/or magnetism \citep{Skumanich1972,Soderblom2010ages,WrightDrake_2011}. Our catalog contains 3561 solar-like variables with identified rotation from the GDR3 table \texttt{vari\_rotation\_modulation} \citep{GaiaDR3_SOS_Rotation} spanning OCs with ages of $6 \lesssim t \lesssim 800$\,Myr. As discussed in \citet{GaiaDR3_variability_classifications}, these variables overlap with the SOLAR\_LIKE and RS~CVn in the supervised classification, cf. also \citet[their Fig.\,7]{GaiaDR3_variability_eyer}. Figure\,\ref{fig:gyrochronoloy} illustrates age trends among these variables using the best rotation periods, $P_\text{rot}$, and the maximum of the percentile-based index in $G$-band, $\mathcal{A}_G$, reported as \texttt{best\_rotation\_period} and \texttt{max\_activity\_index\_g}. Specifically, Fig.\,\ref{fig:gyrochronoloy} shows the per-OC medians of these quantities for OCs with at least 10 such variables with $0.7 \le (G_\text{BP}-G_\text{RP})_0 \le 1.3$\,mag, corresponding to the mass range of $0.8-1.2\,M_\odot$. 
Restricting the color interval to $0.9 \le (G_\text{BP}-G_\text{RP})_0 \le 1.1$\,mag yields a generally consistent picture despite reduced statistics, resulting in a more limited age range of $\log{t} \ge 7.4$. 
Additionally, Fig.\,\ref{fig:gyrochronoloy} shows gyrochronology relations for slowly rotating stars in the middle of our considered color range (assuming $(G_\text{BP}-G_\text{RP})_0=1$~mag) from \cite{AngusAigrain_2015,AngusMorton_2019} and \cite{BoumaPalumbo_2023}. Despite \gaia's individual rotation periods being less precise than \emph{Kepler}'s \citep[e.g.,][]{SantosGodoy-Rivera_2024}, the $\langle P_{\mathrm{rot}}\rangle - \log{t}$ relation reproduces these trends well.
In GDR3, the detection of rotation periods $> 5$\,d is increasingly hampered by \gaia's scanning law \citep[their Fig.~18]{GaiaDR3_SOS_Rotation}. In part due to this selection effect, and in part due to few old clusters in our sample, our catalog does not yield a good measure of the \citet{Skumanich1972} law. However, the catalog's age range is particularly interesting for investigating the earliest ages where gyrochronology relations hold \citep{Boyle2023}. Figure\,\ref{fig:period_color_plots} illustrates period-color diagrams versus age and compares them to the I and C sequences of \citet{Barnes2003}, suggesting that OCs as young as $25-40$\,Myr may be useful to this end. We expect the doubled baseline and improved photometry in GDR4 to significantly improve the inventory of solar-like rotators in OCs in the near future.

The activity index, $\mathcal{A}_G$, also exhibits an age trend, albeit of an unclear functional form, cf. Fig.~\ref{fig:gyrochronoloy}. 
We see that $\mathcal{A}_G$ clearly decreases above $\sim 150-200$\,Myr and exhibits higher levels below, likely due to larger spot size in more magnetically active stars. A hint of saturation may be seen as a leveling off in $\mathcal{A}_G$ at ages $\lesssim 40$\,Myr. However, it is not possible to assign a specific age for saturation, partially due to the large scatter in the data.

\section{Conclusions and outlook}\label{sec:conclusion}

This Letter showcases the power of variable star populations in OCs to create an empirical bird's eye view of stellar evolution. Our high-quality catalog of 34\,760 variable stars residing in 1\,192 OCs is made publicly available. Here, we focused on the relation between the variable star content and ages of OCs. For solar-like rotators, we show that GDR3 activity indicators trace literature relations of gyrochronology well, while providing a useful sample of young clusters near the age where the rotation-activity relation is expected to start. Potential applications of this catalog abound. For example, instability regions may serve as multiple ``standard crayons'' for dereddening \citep[as done for RR~Lyrae stars, e.g., in][]{Nemec1988}, and ages of stellar populations could be determined based on the properties of their variable star types. 

Although the spacecraft was recently decommissioned, \gaia's scientifically most exciting phase is just beginning. GDR4 will provide much improved astrometric OC membership constraints (proper motions to improve by $\sim 1/\sqrt{8}$) as well as much more abundant and detailed variability information derived from a doubled observational baseline and improved processing. Thus, GDR4's high-quality homogeneous all-sky census of populations of variable stars in OCs will uniquely advance our understanding of stellar astrophysics and evolution.

\begin{acknowledgements}
RIA dedicates this work to Gisela L. Anderson. We thank the anonymous referee for constructive reports that improved the manuscript. Useful discussions with L. Eyer are acknowledged. This research received support from the European Research Council (ERC) under the European Union’s Horizon 2020 research and innovation programme (Grant Agreement No. 947660). RIA is funded by the SNSF through a Swiss National Science Foundation Eccellenza Professorial Fellowship (award PCEFP2\_194638). 
This work has made use of data from the European Space Agency (ESA) mission {\it Gaia}.
\end{acknowledgements}


\bibliographystyle{aa} 
\bibliography{biblio.bib} 

\begin{appendix}

\onecolumn
\section{Removed variable stars \label{app:misclassified_variables}}

Of the 34\,760 variable stars in our initial filtered set, we removed five variable stars for the following reasons. The small number of removed stars speaks to the impressive quality of the variable star classification in GDR3 for stars that are relatively nearby -- at least when there are no major complications, such as low S/N data or crowding issues.

Firstly, we removed two stars classified as RR Lyrae variables from our sample (source IDs 5514849058739106048 and 5239771349915805952), both of which reside on the main sequences of very young OCs (10 and 26 Myr). The RR Lyrae SOS \citep{GaiaDR3_SOS_RRLyrae} mode identification (RRc) suggests rather sinusoidal light curves, which are often difficult to classify. Their location in the CaMD and likely sinusoidal light curves suggest that these stars could be eclipsing binaries, although they are not included in the corresponding SOS table. We therefore do not identify any ``young'' RR~Lyrae stars in of the OCs in our sample, despite the recent claim that such stars could form through binary interaction channels at any age and should be frequent within the Solar neighborhood \citep{Bobrick2024}.

Secondly, we considered including additional LPV candidates from the supervised classification \citep{GaiaDR3_variability_classifications} in addition to the SOS LPV catalog \citep{GaiaDR3_SOS_LPVs}. However, we subsequently removed all three candidates based on their under-luminous location in the CaMD. Two of them, GDR3 source IDs 5979495594536212224 and 5979076238279556352, fall near the main sequence and are located in OCs younger than $10\,$\,Myr, indicating a possible nature as YSOs.

\section{Data tables}

\begin{table*}[h]
    \caption{\label{tab:catalog}Catalog of OC members with GDR3 variability classifications.}
    \centering
    \begin{tabular}{@{}ccccccccc@{}}
        \toprule
        GDR3 source ID & 
        OC & 
        Class & 
        SOS CEP & 
        SOS ECL &
        SOS LPV &
        SOS MS &
        SOS Rot. &
        SOS Short \\
        \midrule

\multicolumn{9}{c}{$\cdot \cdot \cdot$} \\ 

414494859812091136 & LISC 3534 & ACV|... & N & N & N & N & N & N \\
1995070793575136256 & NGC 7789 & LPV & N & N & Y & N & N & N \\
2164054838739245824 & NGC 7039 & SPB & N & N & N & N & N & N \\
3158716144317425152 & CWNU 2902 & BCEP & N & N & N & N & N & N \\
5616464656375958784 & NGC 2362 & CV & N & N & N & N & N & N \\
5616530558355526144 & HSC 1902 & ECL & N & Y & N & N & N & Y \\
5934060273758436352 & NGC 6152 & BCEP & N & N & N & N & N & N \\
6056407605465867392 & NGC 4755 & ACYG & N & N & N & N & N & N \\
6056416676437740416 & NGC 4755 & ACYG & N & N & N & N & N & N \\
6059764002146656128 & UBC 290 & CEP & Y & N & N & N & N & N \\

\multicolumn{9}{c}{$\cdot \cdot \cdot$} \\ 
        \bottomrule
    \end{tabular}
    \tablefoot{Shown for ten randomly selected stars only. `SOS' columns are a flag indicating membership in a GDR3 Specific Object Studies table. The full version with 34\,760 stars is available at the CDS.
}
\end{table*}

\begin{table*}[h]
    \caption{\label{tab:catalog-photometry}Catalog of OC members with dereddened absolute photometry.}
    \centering
    \begin{tabular}{ccccccc}
        \toprule
        GDR3 source ID & 
        OC & 
        $M_G$ [mag] & 
        $\left(G_\text{BP} - G_\text{RP} \right)_0$ [mag] &
        $\log \left( t \, \text{[yr]} \right)$\tablefootmark{a} & 
        $A_V$ [mag]\tablefootmark{a} &
        $d$ [pc]\tablefootmark{a} \\
        \midrule

\multicolumn{7}{c}{$\cdot \cdot \cdot$} \\ 

1995019936856791168 & NGC 7789 & 4.38 & 0.78 & 9.19 & 0.78 & 1969 \\
1996807094227640832 & ASCC 128 & 4.51 & 0.82 & 8.04 & 0.32 & 657 \\
3046522836812004352 & FSR 1178 & 4.39 & 0.99 & 6.89 & 0.79 & 1109 \\
3344420600229628032 & NGC 2169 & 6.38 & 2.32 & 6.72 & 0.58 & 935 \\
4092799646329127680 & IC 4725 & 6.99 & 1.34 & 7.98 & 1.36 & 645 \\
4145995805582454400 & HXHWL 66 & -0.26 & -0.21 & 7.32 & 1.44 & 1611 \\
4311542747370036480 & NGC 6709 & 3.82 & 0.66 & 8.21 & 0.84 & 1048 \\
5848893717947216384 & Alessi 6 & 5.71 & 1.06 & 8.56 & 0.67 & 861 \\
6056827996863152512 & NGC 4852 & 1.68 & 0.00 & 8.02 & 1.30 & 1238 \\
6056966123009117312 & OC 0588 & 7.79 & 1.73 & 8.37 & 0.64 & 653 \\

\multicolumn{7}{c}{$\cdot \cdot \cdot$} \\ 
        \bottomrule
    \end{tabular}
    \tablefoot{Shown for ten randomly selected stars only. The full version with 173\,294 stars is available at the CDS. \tablefoottext{a}{Host OC parameters taken from HR24.}
}
\end{table*}

\begin{landscape}
\begin{table*}[h]
    \caption{\label{tab:variables}Summary of variable star classes in OCs and their global properties.}
    \centering
    \begin{tabular}{@{}L{3.5cm}L{4cm}cccccccccc@{}}
        \toprule
        Type & 
        Class name & 
        $n_\text{var}$ & 
        $n_\text{cl}$ & 
        $n_\text{contour}$ & 
        $ \mathcal{F}_{\mathrm{var}}$ &
        $M_{G,0} $ &
        $\left(G_\text{BP} - G_\text{RP} \right)_0$ &
        $ m_{G} $ & 
        $\mathcal{A}$ &
        $\langle\mathcal{A} \rangle$ & 
        $\mathcal{P}$ \\
        & & & & & [\%] & [mag] & [mag] & [mag] & [mmag] & [mmag] & [d] \\

        \midrule

$\alpha$ Cyg & ACYG & 5 & 3 & 5 & 62 & -7.7, -4.0 & -0.5, 0.4 & 5.6, 7.0 & 34, 62 & 52 & - \\
Cepheid & CEP & 7 & 7 & 7 & 50 & -4.6, -2.9 & 0.7, 1.1 & 6.2, 8.6 & 242, 655 & 472 & 5.5, 11.3 \\
Long-period variables & LPV & 9 & 9 & 9 & 15 & -7.1, -1.6 & 1.3, 4.2 & 5.3, 8.9 & 125, 522 & 253 & 128.1, 743.4 \\
Be, $\gamma$ Cas, S Dor., ...\tablefootmark{a} & BE|GCAS|SDOR|WR & 20 & 17 & 18 & 1.6 & -4.0, -0.5 & -0.4, -0.1 & 6.2, 10.9 & 39, 487 & 149 & - \\
$\beta$ Cep & BCEP & 50 & 33 & 43 & 3.0 & -3.8, -0.5 & -0.5, -0.1 & 7.3, 11.6 & 17, 66 & 28 & 0.2, 0.5\tablefootmark{MS} \\
Slowly pulsating B star & SPB & 96 & 72 & 87 & 1.5 & -2.6, 1.4 & -0.3, -0.0 & 4.8, 11.4 & 23, 85 & 43 & 0.08, 4.7\tablefootmark{MS} \\
$\alpha^2$ CVn, Chemically peculiar, ...\tablefootmark{a} & ACV|CP|MCP|ROAM\newline{}|ROAP|SXARI & 355 & 199 & 316 & 1.5 & -2.4, 2.6 & -0.5, 0.3 & 5.4, 10.0 & 12, 104 & 29 & - \\
$\delta$ Sct, $\gamma$ Dor, SX Phe\tablefootmark{a} & DSCT|GDOR|SXPHE & 5407 & 930 & 3678 & 14 & 1.2, 3.4 & -0.0, 0.6 & 8.8, 15.2 & 9, 86 & 23 & 0.04, 9.1\tablefootmark{MS} \\
Eclipsing binaries & ECL & 518 & 327 & 406 & 0.30 & -1.6, 8.1 & -0.4, 1.9 & 7.8, 17.9 & 11, 859 & 182 & 0.2, 55.3 \\
Solar-like variability & SOLAR\_LIKE & 15772 & 795 & 11300 & 22 & 3.9, 7.0 & 0.6, 1.4 & 7.7, 18.0 & 12, 146 & 33 & - \\
RS CVn & RS & 8175 & 851 & 5647 & 9.8 & 4.2, 7.4 & 0.7, 1.6 & 9.2, 18.0 & 39, 308 & 97 & - \\
Rotational Modulation & ROT & 3561 & 375 & - & - & - & - & - & 4, 297 & 50 & 0.3, 44.4 \\
Short-timescale & S & 14 & 13 & 14 & 0.02 & 2.7, 11.1 & -0.2, 2.6 & 14.1, 17.6 & 142, 686 & 311 & - \\
Cataclysmic variables & CV & 4 & 4 & 4 & 0.01 & 4.4, 10.9 & 0.7, 1.7 & 16.7, 17.4 & 497, 3192 & 1585 & - \\
Young stellar object & YSO & 4326 & 169 & 3106 & 32 & 5.3, 10.2 & 1.4, 3.4 & 11.6, 18.0 & 23, 2849 & 116 & - \\
Variable white dwarf & WD & 2 & 1 & 0 & - & 11.0, 11.7 & -0.2, -0.2 & 14.5, 15.2 & 38, 49 & 43 & - \\

        \bottomrule
    \end{tabular}
    \tablefoot{
    $n_\text{var}$ is the total number of variables per class found in all OCs, $n_\text{cl}$ the number of OCs in which they occur, and $n_\text{contour}$ is the number of variable stars in each CaMD contour region identified in Fig.~\ref{fig:variable_hr}. Within these contours, $\mathcal{F}_\mathrm{var}$    is the fraction of stars that correspond to this variable class.
    Brackets (e.g., $\langle m_G \rangle$) denote average  values derived within the contours shown in Fig.\,\ref{fig:variable_hr}; values separated by a comma denote ranges. $m_G$ and $\sigma_{m_G}$ denote the $G-$band apparent (mean) magnitude and its uncertainty, respectively. Script letters $\mathcal{A}$ and $\mathcal{P}$ denote trimmed $G-$band amplitude (\texttt{trimmed\_range\_mag\_g\_fov}) and \gaia{} variability period, respectively. Periods are taken from SOS results when available.
        \tablefoottext{a}{Type contains multiple kinds of associated variable star.}
        \tablefoottext{MS}{Periods taken from the GDR3 SOS main sequence oscillators analysis \citet{GaiaDR3_ms_oscillators}}
}
\end{table*}
\end{landscape}


\onecolumn

\section{Examples of OCs hosting variable stars}\label{app:example_clusters}
\begin{figure*}[h]
\centering
\includegraphics[width=\textwidth]{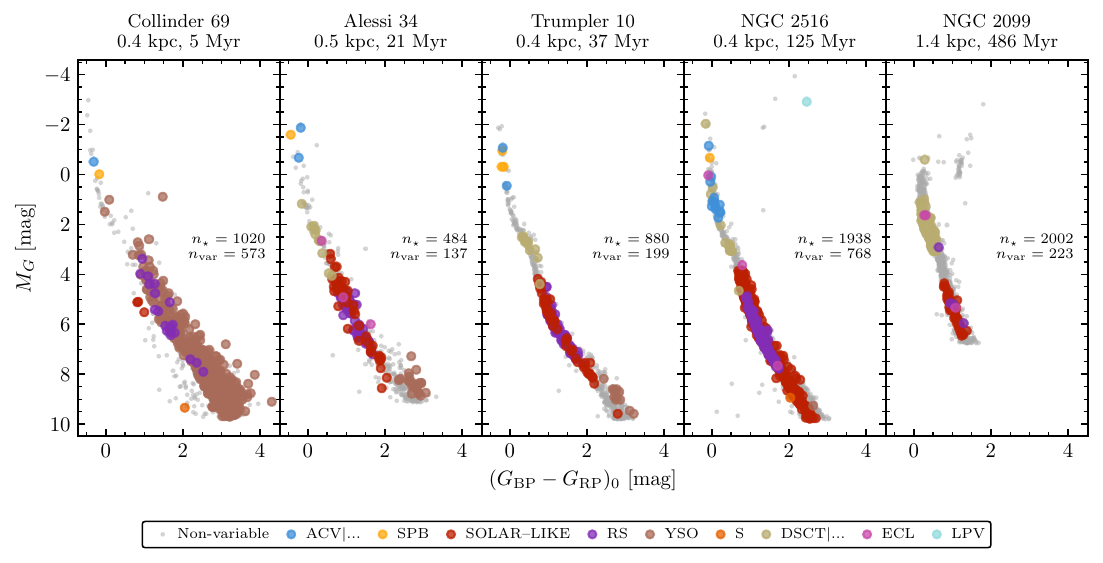}
\caption{Five example OCs with variable star members indicated, with OCs increasing in age from left to right. Non-variable member stars are shown by the small gray dots, with variable member stars shown with colored circles according to the legend at the bottom of the figure. The total number of member stars $n_\star$ and the number that are variable $n_\text{var}$ are indicated in each figure. OCs shown were selected to be both well-populated and to span a useful interval in ages.\label{fig:example_clusters}}
\end{figure*}

\section{On variable star detectability across the catalog}\label{app:detectability}

\begin{figure}
\centering
    \includegraphics[height=2.3in]{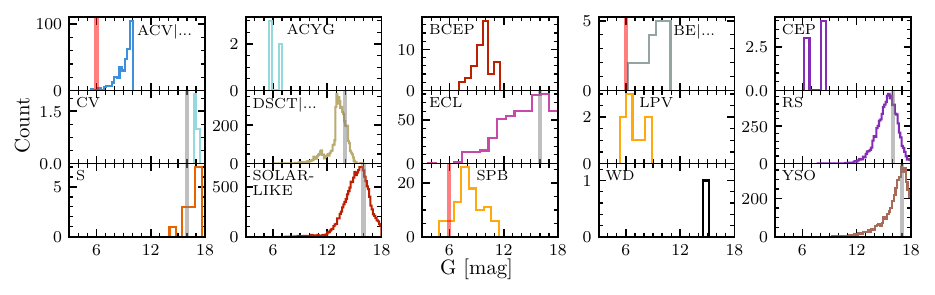}
    \caption{Distribution of $m_G$ among variable stars in OCs per class. Decreasing counts at fainter $m_G$ indicate reduced detectability due to larger photometric errors. Vertical gray bars indicate the limiting magnitudes used to select OC samples for computing $\mathcal{F}$ and in Figs.\,\ref{fig:joy_division} and \ref{fig:variable_member_fraction}, cf. Appendix\,\ref{app:detectability}. Red vertical lines are shown for variable classes where OC samples were limited to avoid saturation effects.}
    \label{fig:detectability}
\end{figure}

\gaia's photometric uncertainties set a detectability floor for variability, cf. Sect.\,\ref{sec:results:localfrac}. Additionally, detectability of variables occurring in specific $M_{G,0}$ ranges depends on $d$ and $A_G$. Hence, variability detection is hampered both among luminous stars reaching $m_G \lesssim 6$\,mag and among stars exceeding $m_G \gtrsim 13$\,mag ($M_{G} \gtrsim 1.5$\,mag at 2\,kpc). While we did not determine overall completeness, we sought to ensure consistent (unknown) completeness levels within each variability class across the OC samples used in Figs.\,\ref{fig:joy_division} and \ref{fig:variable_member_fraction}, and when calculating $\mathcal{F}_\mathrm{var}$. To this end, we did not consider OCs in which the variability contours of Fig.\,\ref{fig:variable_hr} a) for ACV|..., BE|..., and SPBs reached saturation at $m_G < 6$\,mag (OC too close), or b) for CV, DSCT|..., ECL, RS, S, solar-like, and YSOs reached an approximate limiting $m_G$ determined as a decrease in counts with $m_G$, cf. Fig.\,\ref{fig:detectability}. For YSOs and ECLs, we considered smaller ranges in $M_{G,0}$ ($< 8$\,mag and $< 3.5$\,mag, respectively), as labeled in Fig.\,\ref{fig:joy_division} to avoid loss of statistics due to their large $M_{G,0}$ contours.


\section{Alternative variable star CaMDs}\label{app:camd_individual_stars}

\begin{figure*}[ht]
\centering
\includegraphics[width=0.98\textwidth]{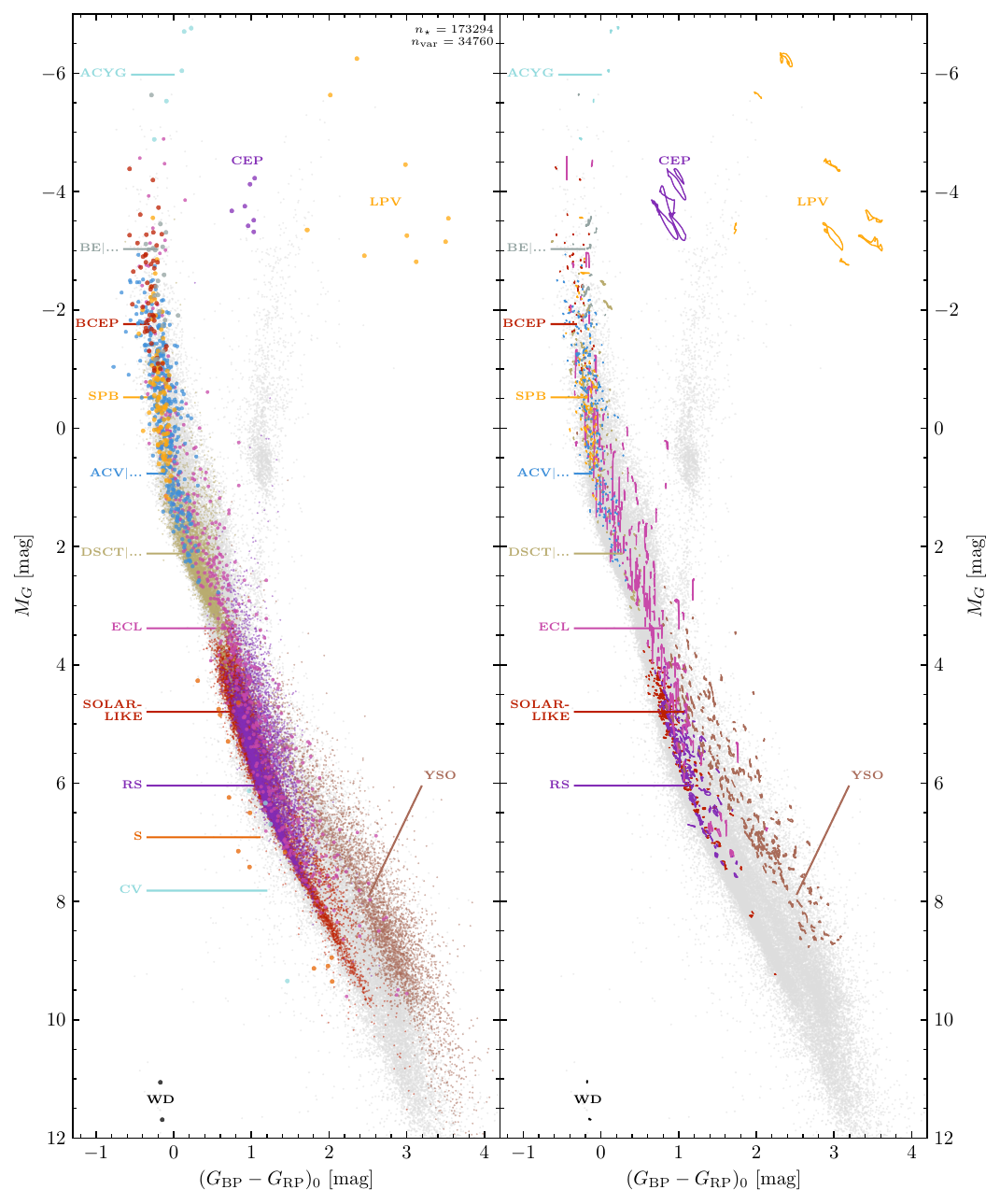}
\caption{\emph{Left:} CaMD with different variability types shown in different colors. Non-variable member stars are shown as small light gray circles. \emph{Right:} Same as left panel, but showing tracks derived from GDR3 epoch photometry, limited to up to 200 variables per class with the highest number of epochs in GDR3 and the lowest chance of having blended transits \citep[following cuts in][Sect. 9.4]{GaiaEDR3_Photometry}. \label{fig:variable_hr_with_points}}
\end{figure*}

\twocolumn 

\section{Fraction of variables with age\label{app:varfrac_age}}

\begin{figure}[h]
\centering
\includegraphics[]{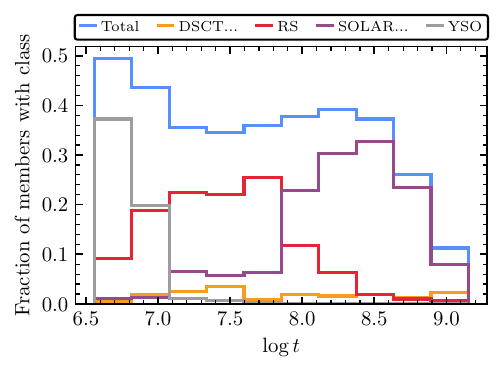}
\caption{Fraction of OC members with assigned variability classes in GDR3 against $\log t$, both in total numbers and for the four most populous variable classes individually. Only stars with $M_{G,0} < 7$\,mag residing in OCs with extinction-corrected distance modulus $\mu + A_G<9$ are shown so that all member stars have $m_G < 16$. These simple cuts aim to improve the similarity of variability detectability with $\log{t}$ compared to not applying cuts, cf. Sect.~\ref{app:detectability}.
\label{fig:variable_member_fraction}}
\end{figure}

\section{Example Period-color plots\label{app:period-color-plots}}

\begin{figure}[h]
\centering
\includegraphics[]{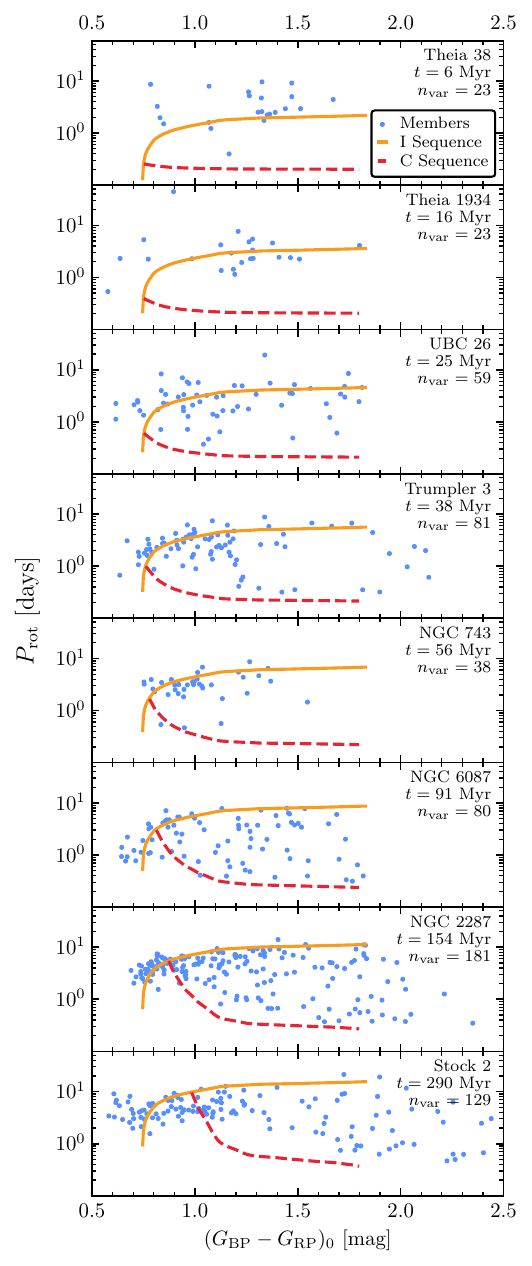}
\caption{Period-color diagrams for three OCs that contain solar-like rotators, shown in descending order of age. Blue points show the GDR3 rotation period of these stars from \citet{GaiaDR3_SOS_Rotation} against dereddened color. The solid orange and dashed red curves show the predicted location of the I and C sequences of slow and fast rotators from \cite{Barnes2003}. The detection of rotation periods $> 5$\,d is increasingly disfavored by the \gaia\ scanning law \citep{GaiaDR3_SOS_Rotation}, which explains the poor match of the $I-$sequence in NGC\,2447. \label{fig:period_color_plots}}
\end{figure}

\end{appendix}

\end{document}